\documentclass[fleqn,10pt]{wlscirep}
\usepackage[utf8]{inputenc}
\usepackage[T1]{fontenc}
\usepackage{lineno}
\usepackage{siunitx}
\usepackage{listings}
\usepackage{tikz}
\usetikzlibrary{positioning, arrows.meta}
\usepackage{ulem}  
\usepackage{glossaries}
\usepackage{csvsimple}
\usepackage{bm} 
\usepackage{todonotes}
\usepackage[version=4]{mhchem}
\usepackage{eurosym}
\usepackage[author={Bobby Xiong}]{pdfcomment}
\usepackage{makecell} 
\usepackage{pdflscape}
\usepackage{array}
\usepackage{hyperref}
\glsdisablehyper 
\let\autocite\cite

\makeatletter
\def\doi#1{%
  \href{https://doi.org/#1}{\textnormal{#1}}%
}
\makeatother

\newacronym{api}{API}{Application Programming Interface}
\newacronym{osm}{OSM}{OpenStreetMap}
\newacronym{entsoe}{ENTSO-E}{European Network of Transmission System Operators for Electricity}
\newacronym[plural=tso,firstplural=Transmission System Operators (TSOs)]{tso}{TSO}{Transmission System Operator}
\newacronym{ac}{AC}{Alternating Current}
\newacronym{dc}{DC}{Direct Current}
\newacronym{rmse}{RMSE}{Root Mean Square Error}
\newacronym{odbl}{ODbL}{Open Data Commons Open Database License}

\newcommand{\colorcode}[1]{\colorbox{gray!20}{\lstinline|#1|}}

\newcolumntype{C}[1]{>{\centering\arraybackslash}p{#1}}

\title{Modelling the high-voltage grid using open data for Europe and beyond}

\author[1,*]{Bobby Xiong}
\author[2]{Davide Fioriti}
\author[1]{Fabian Neumann}
\author[1]{Iegor Riepin}
\author[1]{Tom Brown}
\affil[1]{Technische Universität Berlin, Department of Digital Transformation in Energy Systems (Institute of Energy Technology), Berlin, Germany}
\affil[2]{Università di Pisa, Department of Energy Systems, Territory and Construction Engineering, Pisa, Italy}

\affil[*]{corresponding author: Bobby Xiong (\href{mailto:xiong@tu-berlin.de}{xiong@tu-berlin.de})}

\begin{abstract}
This paper provides the background, methodology and validation for constructing a representation of the European high-voltage grid, including and above 200 kV, based on public data provided by OpenStreetMap. The model-independent grid dataset is published under the Open Data Commons Open Database (ODbL 1.0) licence and can be used for large-scale electricity as well as energy system modelling. The dataset and workflow are provided as part of PyPSA-Eur -- an open-source, sector-coupled optimisation model of the European energy system. By integrating with the codebase for initiatives such as PyPSA-Earth, the value of open and maintainable high-voltage grid data extends to the global context. By accessing the latest data through the the Overpass turbo API, the dataset can be easily reconstructed and updated within minutes. To assess the data quality, this paper further compares the dataset with official statistics and representative model runs using PyPSA-Eur based on different electricity grid representations.
\end{abstract}
\begin{document}

\flushbottom
\maketitle

\thispagestyle{empty}

\section*{Background \& Summary}
\pdfbookmark[1]{Background \& Summary}{background-and-summary}
Energy system models are indispensable tools in today's world in order to understand the complex interactions between energy sources, technologies, policies, and markets. They are used by researchers, industry and policy makers to enable informed decision-making in the transition to a net-zero energy system. However, conclusions drawn from such models are only as good as the underlying data and assumptions. Especially the representation of existing energy infrastructure, such as the electricity grid, can have a deciding impact on future investments derived from such models.\autocite{horschRoleSpatialScale2017} While \glspl{tso} have their own information on the high-voltage grid, this data is often not publicly available to the level of detail needed for academic research purposes. Official institutions like the \gls{entsoe} provide an online map \autocite{entso-eENTSOETransmissionSystem} of the European high-voltage grid. There are however, several limitations typical for these sources: i) there is no underlying, topologically connected dataset, ii) it is not released under an open licence, iii) nor updated frequently, and iv) its geographic detail is limited or highly stylised.

There are previous projects that have modelled the European high-voltage grid or its parts based on \gls{osm} data. Some institutions provide data for particular regions, however all of them come with their individual limitations: While the most trustworthy data comes from \acrshortpl{tso} themselves, they are --- with few regional exceptions \autocite{50hertzStaticGridModel2022} --- not georeferenced \autocite{jaoStaticGridModel2023} or do not cover the entirety of Europe. Datasets from previous academic projects \autocite{egererElectricitySectorData2014,hutcheonUpdatedValidatedPower2013,medjroubiOpenDataPower2017} are either very complex to reproduce or have not been updated for close to a decade. An overview of notable projects and datasets is listed in Table \ref{tab:network_projects}. Proven to be a reliable public data source, we make use of \acrshort{osm} to introduce a transparent workflow in order to create a representation of the European high-voltage grid. On the lack of updates of existing datasets, there are two main advantages of our work compared to previous initiatives. First, our approach uses the \acrshort{osm} Overpass turbo \gls{api}\autocite{raiferOverpassTurbo2024} that always allows to retrieve the latest \acrshort{osm} data. Second, an active \acrshort{osm} community as well as a large user base of PyPSA-Eur and integration into automated workflows mean frequent updates and validation of processed data. Debugging can be easily done with the help of the open source project OpenInfraMap \autocite{garrettOpenInfrastructureMap2024} which renders the \gls{osm} energy infrastructure on an interactive map. Finally, the entire workflow is developed in Python and may hence be more accessible than other implementations which require external dependencies (e.g. SQL databases, commercial software, Java).\autocite{wiegmansGridKitExtractENTSOE2016}

Compared to previous implementations in the global modal PyPSA-Earth,\autocite{parzenPyPSAEarthNewGlobal2023} we significantly improve the work in speed and data quality by taking advantage of the topological, electrical and geographical information available for Europe in \gls{osm}. Given the generic structure of the developed workflow, it can be easily applied to other regions and fed back to the global PyPSA-Earth project. However, the output will directly depend on the \acrshort{osm} data quality for a particular region (e.g. whether data on the substation's geometric footprint is available). To fill in missing data, we introduce cleaning process that yields a representation of the European high-voltage grid. 
We benchmark the processed data against country-level statistics provided by ENTSO-E, concluding that \acrshort{osm} data coverage of the European high-voltage grid is high or even close to complete. These improvements will also contribute to the quality of transmission grids modelled on a global scale in PyPSA-Earth. 
\section*{Methods}
\pdfbookmark[1]{Methods}{methods}
PyPSA-Eur is a spatially and temporally highly resolved, open-source, sector-coupled linear optimisation model that covers the European continent.\autocite{horschPyPSAEurOpenOptimisation2018} The model is build on top of the open-source toolbox PyPSA \autocite{brownPyPSAPythonPower2018} and is suited for operational as well as expansion studies (transmission, generation, and storages). The model includes a stock of existing power plants (processed with the tool powerplanmatching\autocite{gotzensPerformingEnergyModelling2019}) as well as renewable potentials and availability time series (processed with atlite\autocite{hofmannAtliteLightweightPython2021}).
Throughout the last decade, PyPSA-Eur has gained a large user base from academia, industry, and policy makers alike and has been used in a variety of studies.\autocite{neumannPotentialRoleHydrogen2023,victoriaSpeedTechnologicalTransformations2022,brownUltralongdurationEnergyStorage2023,glaumOffshorePowerHydrogen2024,riepinMeansCostsSystemlevel2024,rahdanDistributedPhotovoltaicsProvides2024,grochowiczUsingPowerSystem2024,transnetbwStromnetz2050Studie2022} Other open-source models exist, one notable being OSeMOSYS Global,\autocite{barnesOSeMOSYSGlobalOpensource2022} however it lacks the detailed geographical as well as electrical representation of the transmission grid that PyPSA-Eur provides. With the integration into PyPSA-Eur, we also enable compatibility with additional functions already implemented into the model, such as, but not limited to, the option to enable dynamic line rating\autocite{glaumLeveragingExistingGerman2023} and adding projects under planning (e.g. European Ten-Year Network Development Plan\autocite{entso-eTenYearNetworkDevelopment2020} and the German Network Development Plan\autocite{bnetzaBestaetigungNetzentwicklungsplanStrom2024}).

\begin{figure}[h!]
    \centering
    \resizebox{0.85\textwidth}{!}{
        \begin{tikzpicture}[node distance=1cm, every node/.style={draw, minimum height=1cm, minimum width=0cm, align=center}, >=Latex]
            \node[draw=none] (osm) {\gls{osm}};
            \node[fill=gray!20, font=\itshape] (retrieve) [right=of osm] {retrieve\_osm\_data};
            \node[fill=gray!20, font=\itshape] (clean) [right=of retrieve] {clean\_osm\_data};
            \node[fill=gray!20, font=\itshape] (build) [right=of clean] {build\_osm\_network};
            \node[fill=gray!20, font=\itshape] (base) [right=of build] {base\_network};

            \draw[->] (osm) -- (retrieve) node[midway, above, draw=none, font=\scriptsize, yshift=-2mm] {\textbf{Step 1}};
            \draw[->] (retrieve) -- (clean) node[midway, above, draw=none, font=\scriptsize, yshift=-2mm] {\textbf{Step 2}};
            \draw[->] (clean) -- (build) node[midway, above, draw=none, font=\scriptsize, yshift=-2mm] {\textbf{Step 3}};
            \draw[->] (build) -- (base) node[midway, above, draw=none, font=\scriptsize, yshift=-2mm] {\textbf{Step 4}};
        \end{tikzpicture}
    }
    \caption{Process diagram for creating the European high-voltage grid from \gls{osm} data, representing the snakemake rules.}
    \label{fig:osm_process}
\end{figure}
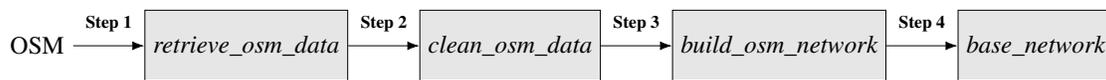

\begin{figure}[!htbp]
    \centering
    \includegraphics{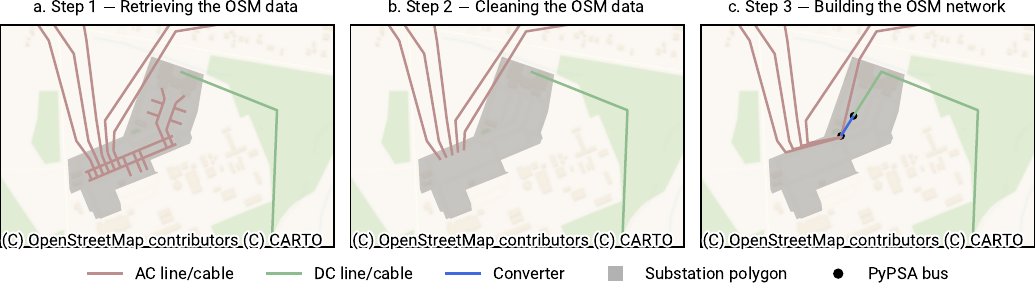}
    \caption{Illustration of the steps to create a PyPSA-ready network from \gls{osm} data. \textit{Note that Step 4 does not make changes to the topology and is hence omitted from this illustration.}}
    \label{fig:all_steps}
\end{figure}

PyPSA-Eur is managed by a workflow management system called Snakemake. \autocite{molderSustainableDataAnalysis2021} Its modular structure enables the addition of new model functionalities and data sources, these can then be toggled using a configuration file. We split the construction of the high-voltage grid into four steps and add them into the existing workflow. We also use the model for validation purposes (see section Technical Validation). While the dataset and its reconstruction is built into PyPSA-Eur, it has the potential to be used in other energy system models, too. To obtain a functioning, topologically connected representation of the European high-voltage grid based on \gls{osm} data, we take the following steps (see Figure \ref{fig:all_steps} for an application to an example region).

\subsection*{Step 1 --- Retrieving the OSM data}
\pdfbookmark[2]{Step 1 - Retrieving the OSM data}{methods-step1}
In \gls{osm}, geographical data is stored in `nodes', `ways', and `relations'. Being the simplest data type, nodes are defined by coordinates and associated parameters. Ways are geometric line strings that connect a set of nodes. Relations can contain nodes, ways, a combination of either or other relations. \autocite{medjroubiOpenDataPower2017} For the purpose of this work, we extract ways for obtaining the outline of substations and \acrshort{ac} power lines and cables. To obtain \acrshort{dc} projects (usually cables, hereafter refered to as \acrshort{dc} links), we use relations. There are two main reasons for this differentiation: i) relations are the more complex data type, coverage for \acrshort{ac} lines and cables is still scarce for Europe, whereas for \acrshort{dc} links, given that there are much fewer of them in Europe, coverage is close to complete (see Table \ref{tab:dc_links}); ii) Some \acrshort{dc} links contain multiple components, accessing relations allows us to efficiently aggregate and simplify the links for the purpose of static energy system modelling.
First, we retrieve the raw data using the Overpass turbo \gls{api}\autocite{raiferOverpassTurbo2024} (Figure \ref{fig:all_steps}.a). Specifically, we query the \gls{osm} database for electricity grid related features (see Table \ref{tab:overpass_queries}). Note that Overpass turbo has a limit on the size and number of requests for each query and provides the \gls{api} under a fair use policy. To avoid unnecessary load and re-use of data, we provide a prepared transmission grid for for download via Zenodo. \autocite{xiongPrebuiltElectricityNetwork2024}

\subsection*{Step 2 --- Cleaning the OSM data}
\pdfbookmark[2]{Step 2 - Cleaning the OSM data}{methods-step2}
While \gls{osm} provides a rich dataset, it is not directly usable for energy system modelling. Next to geospatial coordinates, \gls{osm} includes tags that provide feature-specific additional information. Depending on the individual feature, data may however contain noise, be incomplete, or inconsistent. After importing the retrieved raw data into a pandas dataframe,\autocite{mckinneyDataStructuresStatistical2010} we apply a series of steps including heuristics to clean and fill in the missing information (see example in Table \ref{tab:aclines_example1}). We then use the power of geopandas\autocite{jordahlGeopandasGeopandasV02020} to perform geospatial operations (including but not limited to spatial joins, intersections, buffering, etc.).

\subsubsection*{Substations and transformers}
To obtain the set of substations, we filter for substations with a voltage level within the scope of interest, i.e. between \acrshort{ac} \SI{200}{\kilo\volt} and \SI{750}{\kilo\volt}. Where available, we extract the polygon shape of the substations, stored in the element's geometry. This allows us to differentiate between internal and external grid components. 
Note that information on transformers are not extracted from \acrshort{osm}, as i) we cannot adequately evaluate their coverage and ii) this data is not sufficient to create a topologically connected network. Instead, we use a needs-based approach, i.e. adding a single transformer of \SI{2000}{\mega\watt} between buses of different voltages within the perimeter of the same substation. This is in line with previous approaches to obtain the ENTSO-E map based transmission grid.\autocite{horschPyPSAEurOpenOptimisation2018}

\subsubsection*{AC lines and cables} In a first step, we clean the tag columns to only contain the correct data type and unit, as shown in Table \ref{tab:aclines_params} for \acrshort{ac} power lines and cables. The minimum parameters that need to be given for a particular line or cable are `voltage' (in \si{\volt}) and `power' (string: `line' or `cable').
\begin{table}[ht]
    \centering
    \begin{tabular}{|l|c|c|}
    \hline
    \textbf{Tag} & \textbf{Data type} & \textbf{Example} \\
    \hline
    cables & numeric & 9 \\
    \hline
    circuits & numeric & 3 \\
    \hline
    frequency & numeric & \SI{50}{\hertz} \\
    \hline
    power & string & line \\
    \hline
    voltage & numeric & \SI{380000}{\volt} \\
    \hline
    \end{tabular}
    \caption{Key tags/parameters for \acrshort{ac} power lines and cables.}
    \label{tab:aclines_params} 
\end{table}

We filter for the entries with a voltage level including and above \acrshort{ac} \SI{200}{\kilo\volt}. While data for the mid- to low-voltage grid is also partially available in \gls{osm}, public statistics are scarce, making validation of such data difficult. As not all entries contain clean or complete data, we make heuristic assumptions to fill in the gaps, as illustrated in Tables \ref{tab:aclines_example1} and \ref{tab:aclines_example2}. 

\begin{table}[!htbp]
    \centering
    \begin{tabular}{|l|c|c|c|c|c|c|}
    \hline
    \textbf{line id} & \textbf{cables} & \textbf{circuits} & \textbf{frequency} (Hz) & \textbf{type} & \textbf{voltage}  (V) \\
    \hline
    way/1 &  & 2 & 50 & cable & \SI{380000}{} \\
    \hline
    way/2 & 3 &  & 50 & cable & \SI{380000}{} \\
    \hline
    way/3 & 9 & 1;2 & 50 & line & \SI{380000}{}; \SI{220000}{} \\
    \hline
    way/4 & 9 & 3 & 50; 50 & line & \SI{380000}{}; \SI{220000}{} \\
    \hline
    way/5 & 8 &  & 50 & line & \SI{110000}{}; \SI{220000}{} \\
    \hline
    way/6 &  &  & 50 & cable & \SI{300000}{}\\
    \hline
    \end{tabular}
    \caption{Illustrative example of \acrshort{ac} lines and cables input data.}
    \label{tab:aclines_example1}
\end{table}

\begin{table}[!htbp]
    \centering
    \begin{tabular}{|l|c|c|c|c|c|}
    \hline
    \textbf{line id} & \textbf{circuits} & \textbf{frequency (Hz)} & \textbf{type} & \textbf{voltage (V)} \\
    \hline
    way/1 & 2 & 50 & cable & \SI{380000}{} \\
    \hline
    way/2 & \cellcolor{yellow!50}1 & 50 & cable & \SI{380000}{} \\
    \hline
    way/3-1 & \cellcolor{yellow!50}1 & 50 & line & \cellcolor{yellow!50}\SI{380000}{} \\
    \hline
    way/3-2 & \cellcolor{yellow!50}2 & 50 & line & \cellcolor{yellow!50}\SI{220000}{} \\
    \hline
    way/4-1 & \cellcolor{yellow!50}1 & \cellcolor{yellow!50}50 & line & \cellcolor{yellow!50}\SI{380000}{} \\
    \hline
    way/4-2 & \cellcolor{yellow!50}1 & \cellcolor{yellow!50}50 & line & \cellcolor{yellow!50}\SI{220000}{} \\
    \hline
    \sout{way/5-1} & \cellcolor{yellow!50}\sout{1} & \sout{50} & \sout{line} & \cellcolor{yellow!50}\sout{\SI{110000}{}} \\
    \hline
    way/5-2 & \cellcolor{yellow!50}1 & 50 & line & \cellcolor{yellow!50}\SI{220000}{} \\
    \hline
    way/6 & \cellcolor{yellow!50}1 & 50 & cable & \SI{300000}{}\\
    \hline
    \end{tabular}
    \caption{Illustrative example of \acrshort{ac} lines and cables after cleaning. Changes highlighted in yellow.}
    \label{tab:aclines_example2} 
\end{table}

For each line or cable we use the most specific information available that the data provides. In a three-phase \acrshort{ac} high-voltage system, we assume three cables to form an \acrshort{ac} circuit (e.g. way/2).\autocite{kirschenPowerSystemsFundamental2024} If a way contains multiple data points split by semicolons (i.e. transmission lines sharing overhead line routes), we split the entries into individual lines, accordingly. In this process, we preserve the original \gls{osm} identifier and its associated geometries. We add a numbered suffix after the split to maintain unique line ids (e.g. way/3 becomes way/3-1 and way/3-2). The given electric parameters are mapped according to the semicolon splits. In some cases however, where the number of data points across columns is not equal (e.g. way/4 and way/5), we make the following assumption: we take the floor of the number of circuits divided by the number of entries in the voltage column. In the absence of better information, this may lead to an underestimation of the real number of circuits. If no information on cables nor circuits are available, we assume a single circuit, provided that a voltage level is given (e.g. way/6). Finally, we remove all ways which represent bus bars and lines which are located fully inside of a substation outline (Figure \ref{fig:all_steps}.b), as they are considered internal elements of the substation and provide no additional information for the purpose of static analyses.
\subsubsection*{DC links and converters}
Due to their distinct electrical properties, we treat \acrshort{dc} links differently from \acrshort{ac} lines and cables. To avoid double counting, we remove all \acrshort{dc} links from the original way queries. Instead, we query the \gls{osm} database for relations that contain \acrshort{dc} links. As data on \acrshort{dc} projects are widely available and because there are fewer of them,\autocite{entso-eENTSOETransmissionSystem,pierriChallengesOpportunitiesEuropean2017} we contribute to the \gls{osm} database by adding missing parameters such as the nominal rating and voltage level. This signifies the ease of data improvements with OSM for the benefit of all. In order for future \acrshort{dc} projects to be traced by our workflow, the following tags are required: `route' = `power', `frequency' = 0, and `rating' in `X MW' format. \acrshort{dc} components need to be correctly linked in the \acrshort{osm} database as member (either `cable' or `line') of the parent relation, respectively.

\begin{figure}[!htbp]
    \centering
    \includegraphics{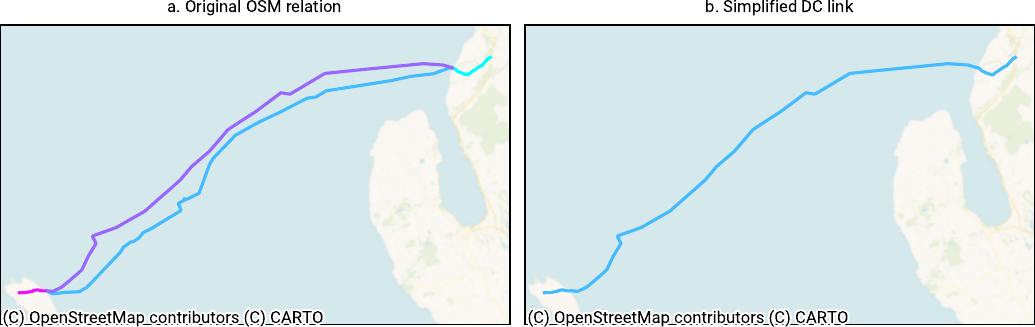}
    \caption{Example -- Simplification of a \acrshort{dc} link relation: Moyle interconnector (Scotland - Northern Ireland), \acrshort{osm} relation ID \href{https://www.openstreetmap.org/relation/6914309}{6914309}. \textit{Different colors show the four segments, i.e. ways that compose the \acrshort{osm} relation.}}
    \label{fig:dc_example}
\end{figure}

Note some relations contain multiple \acrshort{dc} link segments or components (e.g. converter stations or grounding), these are simplified into a single line with the sum of their nominal ratings. Figure \ref{fig:dc_example} shows an example of how the Moyle interconnector from Northern Ireland to Scotland (Figure \ref{fig:dc_example}.a) is simplified (Figure \ref{fig:dc_example}.b). In this simplification, we preserve original end points of the \acrshort{dc} link and the longest connected path. In PyPSA, converters are modelled as links connecting two buses. In analogy to how transformers are introduced to the network ex-post. i.e. connecting the terminals of the \acrshort{dc} link, where the converter stations are also located on \acrshort{osm} and the closest \acrshort{ac} bus in the transmission grid. As such, we guarantee that \acrshort{dc} links are always topologically connected.

\subsection*{Step 3 --- Building the OSM network}
\pdfbookmark[2]{Step 3 - Building the OSM network}{methods-step3}
\begin{figure}[!htbp]
    \centering
    \includegraphics[width=\linewidth]{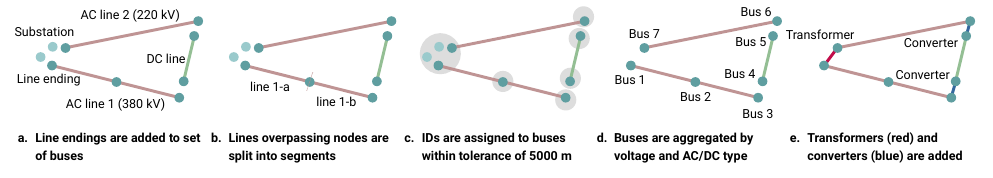}
    \caption{Illustration of building a topologically connected grid.}
    \label{fig:build_osm_network}
\end{figure}

After having obtained a cleaned dataset, we need to ensure that the grid components are topologically and electrically connected. For this purpose, we propose a complete graph-based structure, i.e. buses/substations are connected by \acrshort{ac} and \acrshort{dc} lines, as well as converters and transformers. We introduce the following procedure, illustrated in Figure \ref{fig:build_osm_network}.

First, we add the endpoints of both \acrshort{ac} and \acrshort{dc} lines to the set of buses (Figure \ref{fig:build_osm_network}.a). This step ensures that each line has two corresponding buses to which it is initially connected. Note that this may introduce potential duplicates of buses beyond those already present --- which we will take care of later. Next, we take of long lines that are overpassing (multiple) substations. Here we assume that in reality, the substation is electrically connected to the overpassing line. To achieve this, we split the original line into the subsegments between the intersecting nodes and add a suffix to its original identifier (Figure \ref{fig:build_osm_network}.b), to i) preserve the uniqueness of each line and ii) still allow for tracing back the line on \acrshort{osm} or OpenInfraMap.\autocite{garrettOpenInfrastructureMap2024} We update the endpoints of the split lines, accordingly. Third, we cluster all buses within a radius of \SI{5000}{\meter}. We do this for several reasons, i) to avoid duplicates of buses that represent the same substation in reality, ii) to improve the computational efficiency of the model and iii) to improve the topological connectedness of the obtained network. In an interative procedure, buses within the given radius are assigned the same ID (Figure \ref{fig:build_osm_network}.c). Fourth, we cluster all buses assigned to the same station ID and \acrshort{ac}/\acrshort{dc} type. For each of the clusters, a bus is created. Connected lines are remapped, accordingly (Figure \ref{fig:build_osm_network}.d). Finally, we ensure that all given components are correctly electrically connected, i.e. \acrshort{ac} buses of different voltage levels at the same substation are linked by transformers, \acrshort{dc} buses which represent endpoints of \acrshort{dc} lines are connected to the closest \acrshort{ac} bus through converters (Figure \ref{fig:build_osm_network}.e) of equal nominal rating, respectively.

The proposed methodology is an enhanced version of the data processing integrated in the PyPSA-Earth model \autocite{parzenPyPSAEarthNewGlobal2023} which has previously demonstrated the potential of using \acrshort{osm} data for global energy system modelling. Our proposed methodology improves the original implementation in efficiency, computational performance and enhancing the representation of the European high-voltage grid (see Technical Validation Section).

\subsection*{Step 4 --- Creating a PyPSA-ready base network}
\pdfbookmark[2]{Step 4 -Creating a PyPSA-ready base network}{methods-step4}
In a final step, we create a network ready for modelling within PyPSA and PyPSA-Eur. The benefit of our our dataset is that it is provided in .csv format and easily readable. Hence, it may potentially be used outside our given use case, directly or with small adaptations. 

Based on the voltage levels, we map each line to a standard line type library provided with PyPSA \autocite{brownPyPSAPythonPower2018,horschPyPSAEurOpenOptimisation2018,oedingElektrischeKraftwerkeUnd2016,thurnerPandapowerOpenSourcePython2018} with the closest voltage (e.g. \SI{400}{\kilo\volt} is mapped to a \SI{380}{\kilo\volt} line type). Using the standard grid model provided by 50Hertz, \autocite{50hertzStaticGridModel2022} we demonstrate in Figure \ref{fig:scatter_reactance_resistance} that this approach is effective for this particular region in the absence of more accurate data. Using the geometry line length and number of circuits, we calculate the electric parameters, such as impedance, reactance and apparent power $S_{nom}^{AC}$ (Eq. \ref{eq:s_nom}), as these are not contained in the original \gls{osm} input data. We apply a factor of 0.7 to approximate the N-1 security margin (Eq. \ref{eq:s_n-1}). \autocite{shokrigazafroudiTopologybasedApproximationsContingency2022,horschPyPSAEurOpenOptimisation2018} Note that this factor can be individually set in the configuration file of PyPSA-Eur. We provide an overview of all resulting \acrshort{ac} lines and cables in Table \ref{tab:s_nom}.

\begin{align}
    S_{nom}^{AC} &= n_{circuits}\cdot\sqrt{3}\cdot U_{nom}^{OSM} \cdot I_{nom}^
    {pandapower} \label{eq:s_nom} \\
    S_{n-1}^{AC} &= 0.7 \cdot S_{nom}^{AC} \label{eq:s_n-1}
\end{align}

For \acrshort{dc} links, we use the provided length and nominal rating, directly. We assume that all \acrshort{dc} links can be operated in both directions. Lastly, after transformers and converters have been added to the network, we remove all unconnected or islanded components. 

\section*{Data Records}
\begin{figure}[!htbp]
    \centering
     \includegraphics[width=0.99\textwidth] {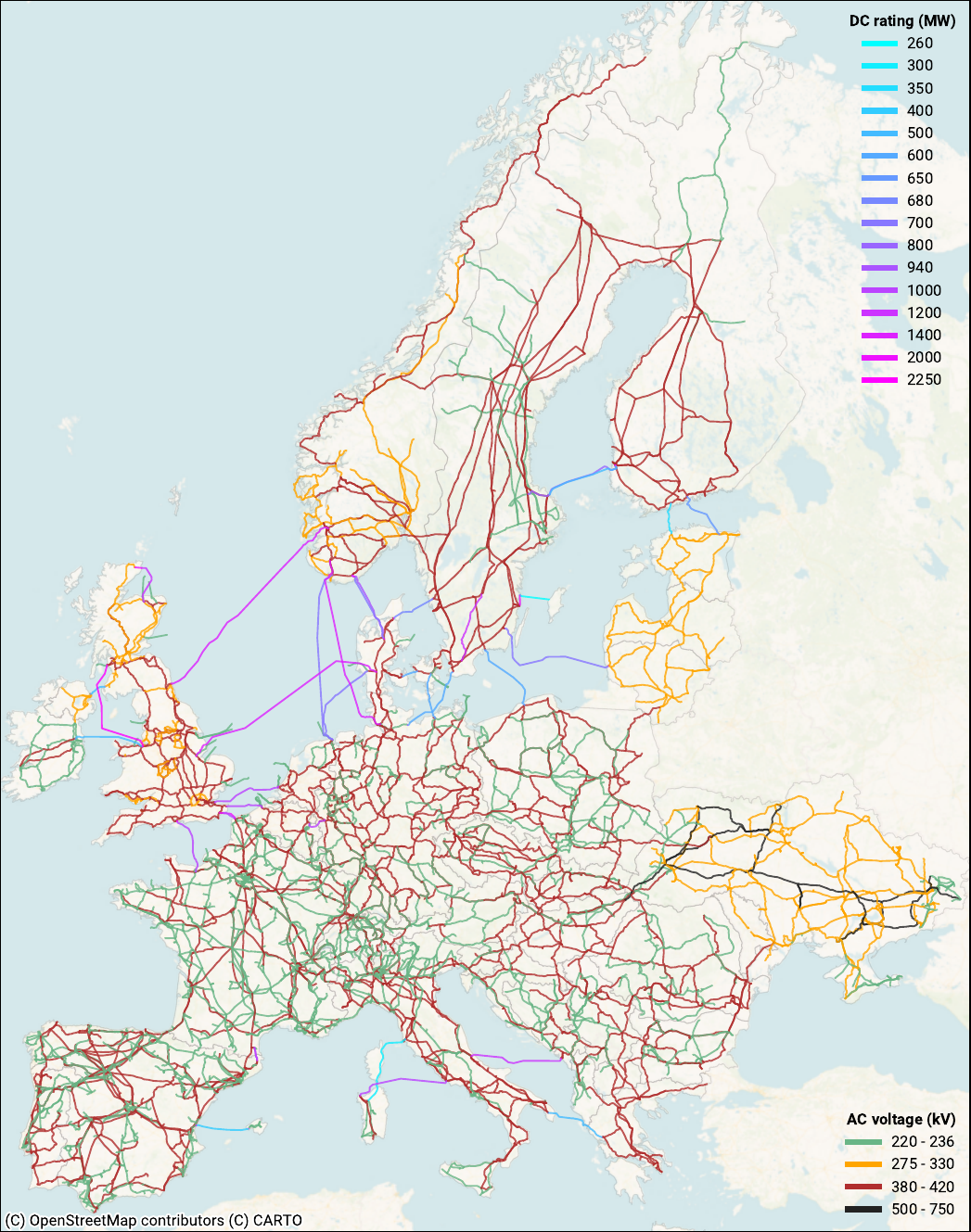}
    \caption{Map of the \acrshort{osm}-based European high-voltage grid. \textit{This map was generated using the grid dataset provided with this publication. \autocite{xiongPrebuiltElectricityNetwork2024}}}
    \label{fig:osm_map}
\end{figure}

\pdfbookmark[1]{Data Records}{data-records}
The compiled representation of the European high-voltage grid (Figure \ref{fig:osm_map}) is hosted online in .csv format and can be downloaded via the Zenodo repository.\autocite{xiongPrebuiltElectricityNetwork2024} The dataset includes the geographical scope of the ENTSO-E member states (without Cyprus, Iceland, Kosovo, and Turkey). 
We continuously update the dataset as the underlying \gls{osm} input data and the workflow is improved. As of the submission of this paper, the resulting network (Figure \ref{fig:osm_map}) contains 5848 buses, 7320 \acrshort{ac} lines and cables, 36 simplified/aggregated \acrshort{dc} links (and converters at their endpoints, respectively), and 1059 \acrshort{ac} transformers, comprising a total of \SI{261757}{\kilo\meter} in total route length.

\newpage
\newpage
\section*{Technical Validation}
\begin{figure}[!htbp]
    \centering
    \includegraphics{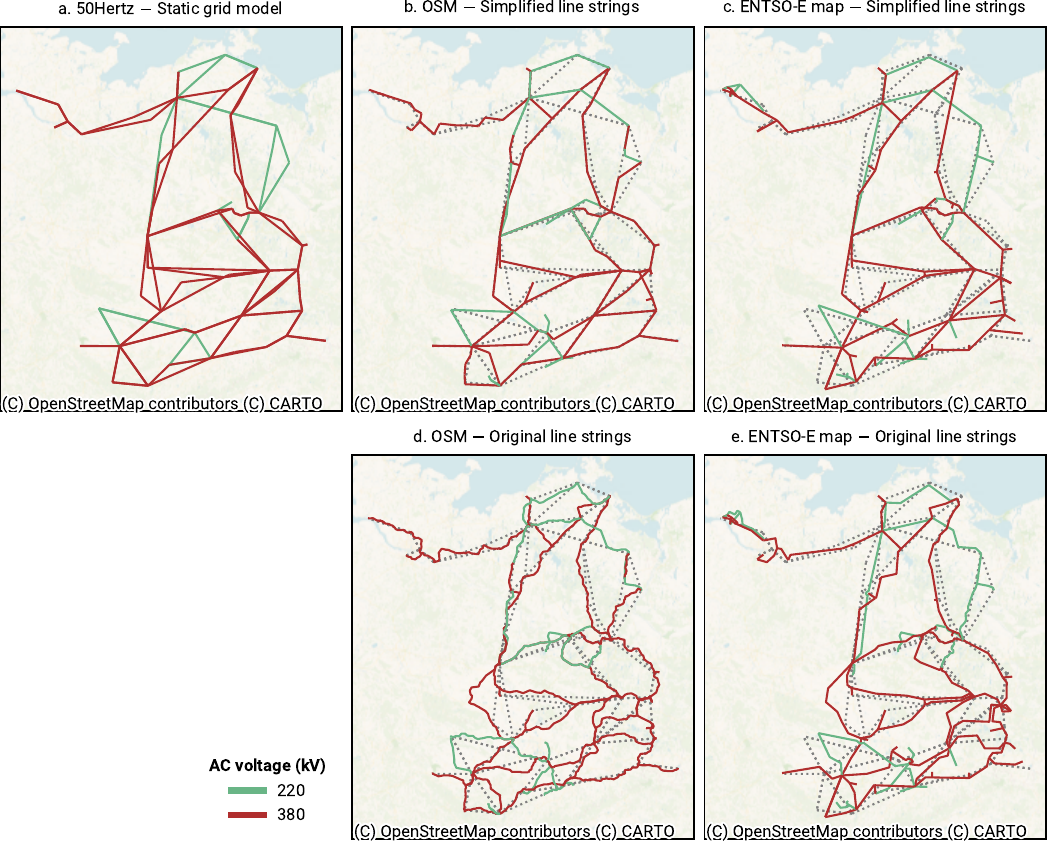}
    \caption{Comparison of \acrshort{osm} and \acrshort{entsoe} map-based transmission grid with reference 50Hertz static grid model.\autocite{50hertzStaticGridModel2022} \textit{Dashed grey lines underneath show the 50Hertz static grid model for comparative purposes. Note that the geospatial data is provided in simplified, point-to-point form, only. Many of the dashed grey lines are also included in the \acrshort{osm} or \acrshort{entsoe} map-based grid representations, however, intermediary buses exists along the line. As such, in their simplification they are not simplified to the level of the 50Hertz static grid model.}}
    \label{fig:osm_vs_50Hertz_sgm}
\end{figure}

\pdfbookmark[1]{Technical Validation}{technical-validation}
We perform two validation steps for assessing the quality of the dataset. First, we compare the dataset with official inventory statistics provided by \acrshort{entsoe}. Second, we compare the results of an representative PyPSA-Eur model instance based on the two high-voltage grid datasets: OSM (presented in \ref{fig:osm_map}) and an extract from the online ENTSO-E map using GridKit tool (referred to as ‘ENTSO-E map’).\autocite{wiegmansGridKitExtractENTSOE2016} This network is currently being used by numerous PyPSA-Eur users, and is hence a good reference for comparison. In Figure \ref{fig:osm_vs_50Hertz_sgm}, we further provide a comparison between a geo-referenced, official dataset by 50Hertz\autocite{50hertzStaticGridModel2022} and the \acrshort{osm}-based grid of the region, respectively.

\subsection*{Comparison with ENTSO-E statistics and map}
\begin{figure}[!htbp]
    \centering
    \includegraphics{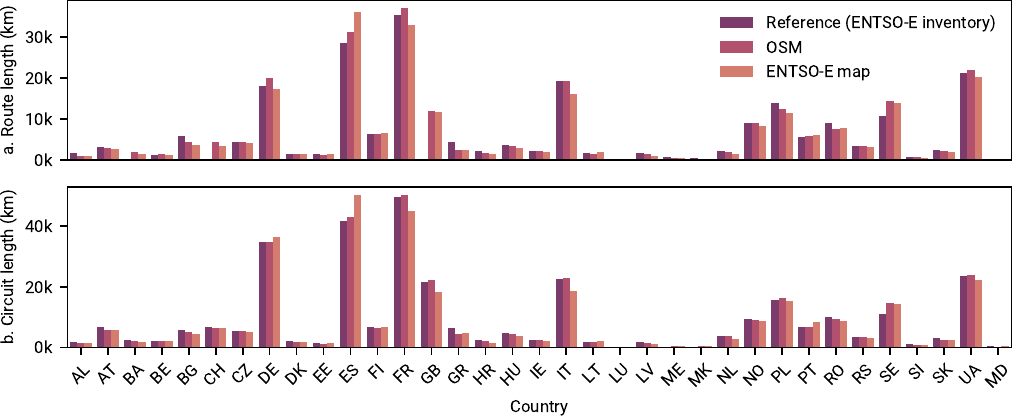}
    \caption{Comparison of total route and circuit lengths per country.}
    \label{fig:bar_lengths}
\end{figure}

\pdfbookmark[2]{Comparison with ENTSO-E statistics and map}{technical-validation-entsoe}
Based on \acrshort{entsoe}'s 2023 inventory of transmission,\autocite{entso-eInventoryTransmission20232024} we first compare the total route (a) and circuit lengths (b) of \acrshort{ac} lines and cables on a per country level (Figure \ref{fig:bar_lengths}.). Note that the inventory does not include all statistics for each country, i.e. route lengths are missing for Bosnia and Herzegovina, Switzerland, and Great Britain, while circuit lengths are missing for Montenegro and North Macedonia. While Ukraine and the Republic of Moldova have joined \acrshort{entsoe} on 1 January 2024 and 22 November 2023 as full and observing members, respectively, their inventory are not yet included in the dataset. For these two countries, we take reference data from third party sources.\autocite{cigrePowerSystemUkraine2018,globaldataTopFiveTransmission2023,moldelectricaTechnicalEconomicIndicators2023}

We find that our transmission grid based on \gls{osm} data is in agreement with the \acrshort{entsoe} inventory. Calculating the Pearson correlation coeffficient for both route and circuit lengths between the official statistics and the respective transmission grid representations, we see an overall improvement from the \acrshort{entsoe} map ($\rho_{routes} = 0.9497$ and $\rho_{circuits} = 0.9862$) to \gls{osm} ($\rho_{routes} = 0.9636$ and $\rho_{circuits} = 0.9980$) in the reproduction of the high-voltage grid (\SI{220}{\kilo\volt} to \SI{750}{\kilo\volt}). One of the key reasons for these improvements is the much higher level of geographic detail of lines and cables in the \gls{osm}-based transmission grid compared to the stylised lines on \acrshort{entsoe}'s interactive map. We observe larger discrepancies for Sweden, where both transmission grid representations seem to overestimate the total lengths of the inventory.

\begin{figure}[!htbp]
    \centering
    \includegraphics{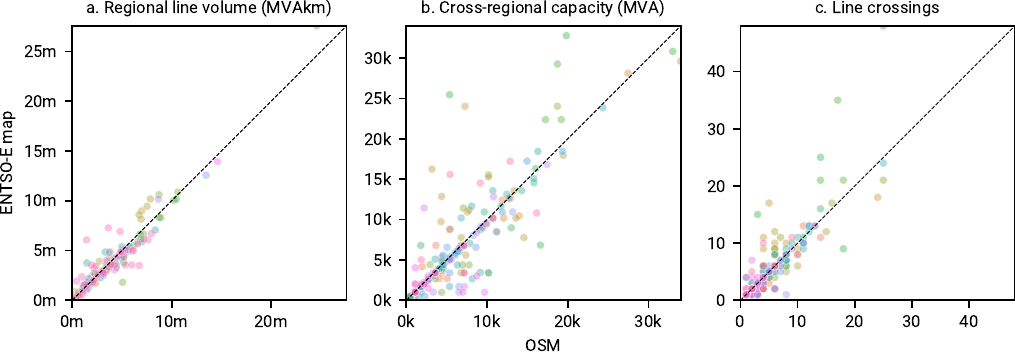}
    \caption{Comparison of line volume per NUTS1 region $-$ Colors represent individual countries. \textit{Line volume is the product of the nominal capacity and the length, summed over all lines within the region.}}
    \label{fig:scatter_joined}
\end{figure}

As another validation step, we compare the total line volume on NUTS1 region level (Figure \ref{fig:scatter_joined}.a). Here we see strong similarities between the \acrshort{osm}-based and the extracted transmission grid of the \acrshort{entsoe} map, with few outliers ($\rho_{MVAkm} = 0.9484$). While a higher geospatial resolution of lines of the \gls{osm}-based transmission grid may contribute to the increase in line volume on average, we calculate $S_{nom}^{AC}$ using the more differentiated voltage levels given in the \gls{osm} data (see Table \ref{tab:s_nom}) as opposed to the clustered voltage levels given in the \acrshort{entsoe} map, i.e. \SI{220}{\kilo\volt}, 300 to \SI{330}{\kilo\volt}, 380 to \SI{400}{\kilo\volt}, \SI{500}{\kilo\volt}, and \SI{750}{\kilo\volt}. This may lead to a more accurate representation of the line volume in the \gls{osm}-based transmission grid.

To assess the transmission capacity across regions, we compare the capacity (Figure \ref{fig:scatter_joined}.b) and number of line crossings (Figure \ref{fig:scatter_joined}.c) per NUTS1 border (Figure \ref{fig:scatter_joined}.b). While the two transmission grids strongly correlate, we observe notable differences at individual borders ($\rho_{MVA} = 0.8491$), the same is true for the absolute number of line crossings ($\rho_{crossings} = 0.8573$). Due to different quality in geospatial information contained in both transmission grid representations, buses in one may be offset (or not even exist) in the other. Stronger outliers can primarily be traced back to buses close to NUTS1 borders (Figure \ref{fig:scatter_joined}.b). Notable outliers are located in Spain (light green), in Ukraine (pink), and southwestern parts of Germany (ocher). The reasons for discrepancies can be manifold, e.g. due to differences in exact locations of boundary nodes, missing, outdated or wrong data. In the case of Spain, the main reason for the large discrepancy lies in the representation of the high-voltage grid in and around the Madrid area, where the \acrshort{entsoe} map is stylised for clarity. A comparison with an official map provided by the National Geographic Institute of Spain\autocite{institutogeograficonacionalEnergiaMapaRed2016} confirms the \gls{osm} topology to be more accurate.

\subsection*{Comparison of model results}
\pdfbookmark[2]{Comparison of model results}{technical-validation-model-results}
In order to assess the impact of the new transmission grid representation, we compare the results of a representative PyPSA-Eur model run based on the \gls{osm} dataset to a run based on the \acrshort{entsoe} map which is currently being used in PyPSA-Eur. \autocite{horschPyPSAEurOpenOptimisation2018} Note that the results shown in this publication are based on version 0.3 of the released prebuilt high-voltage grid representation on Zenodo.\autocite{xiongPrebuiltElectricityNetwork2024} We use the same model setup and input data in both model runs, except for the grid representation. We focus on the electricity sector, taking techno-economic assumptions projected for the year 2030. We allow for capacity expansion in renewable energy as well as gas-fired generation capacities. To narrow down the effect of the transmission grid, we do not allow for grid expansion and disable dynamic line rating. We set the carbon price to 100 \euro{} per tonne of \ce{CO2} emitted.

\begin{figure}[!htbp]
    \centering
    \includegraphics{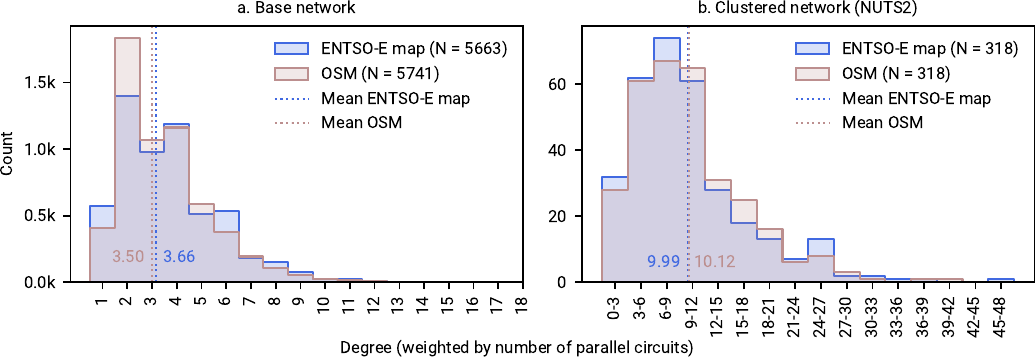}
    \caption{Comparison of the weighted degree distribution in both transmission grid representations before and after clustering (NUTS2). \textit{Ukraine at geoBoundaries\autocite{runfolaGeoBoundariesGlobalDatabase2020} administration level 1, Moldova in full bus resolution. A comparison at NUTS3 resolution is provided in Figure \ref{fig:hist_degree_nuts3}.}}
    \label{fig:hist_degree_nuts2}
\end{figure}

As the number and exact locations of buses differ, we cluster the networks to make them comparable (N = 318 regions/buses). In energy system modelling, clustering is often motivated by the spatial reduction of the optimisation problem. While oftentimes clustering algorithms based on grid topology or resource class are used, \autocite{frysztackiComparisonClusteringMethods2022} we are interested in the regional differences of the two grid representations. As such, we map the buses of both grids to clusters based on administrative boundaries, i.e. NUTS2. For non-NUTS countries such as Ukraine, we use the administration level 1 (geoBoundaries\autocite{runfolaGeoBoundariesGlobalDatabase2020}) and for Moldova we keep the full high-voltage substation resolution, i.e. 8 nodes. This yields 318 regions/buses for each of the clustered high-voltage grids, respectively. Note that the clustering process in PyPSA-Eur involves a transformation of all transmission lines and cables to the default voltage level of \SI{380}{\kilo\volt}. We then run the model at hourly resolution of the year 2030, yielding 8760 time steps. 

Figure \ref{fig:hist_degree_nuts2} compares the weighted degree distribution for the two network topologies. We weight the degree by the number of parallel circuits (Eq. \ref{eq:weighted_degree}) to account for potential different representations of lines and links connecting the same two buses (e.g., single lines with multiple number of circuits or multiple lines with single circuits). If \( G = (V, E) \) is a weighted graph with vertex set \( V \) (buses) and edge set \( E \) (lines, links, converters, and transformers), and each edge \( e \in E \) has a weight \( w(e) \), then the weighted degree of a vertex \( v \in V \) is given by:

\begin{equation}
    d_w(v) = \sum_{e \in \text{IncidentEdges}(v)} w(e)
    \label{eq:weighted_degree}
\end{equation}

where \( \text{IncidentEdges}(v) \) is the set of edges incident to \( v \). We find that the two base networks have a similar weighted degree distribution (Pearson correlation coefficient $\rho_{degree,base} = 0.8518$). Notably, the \acrshort{osm}-based transmission grid demonstrates a higher number of buses with degree 2. Clustering the two networks before running the optimisation problem will increase the Pearson correlation coefficient to $\rho_{clustered,base} = 0.8769$ (using unbinned data), $\rho_{clustered,base} = 0.9937$ (using binned data) --- indicating that at NUTS2 resolution, the two networks are very similar in terms of connectivity and adequately represent the real grid.

\begin{figure}[!htbp]
    \centering
    \includegraphics{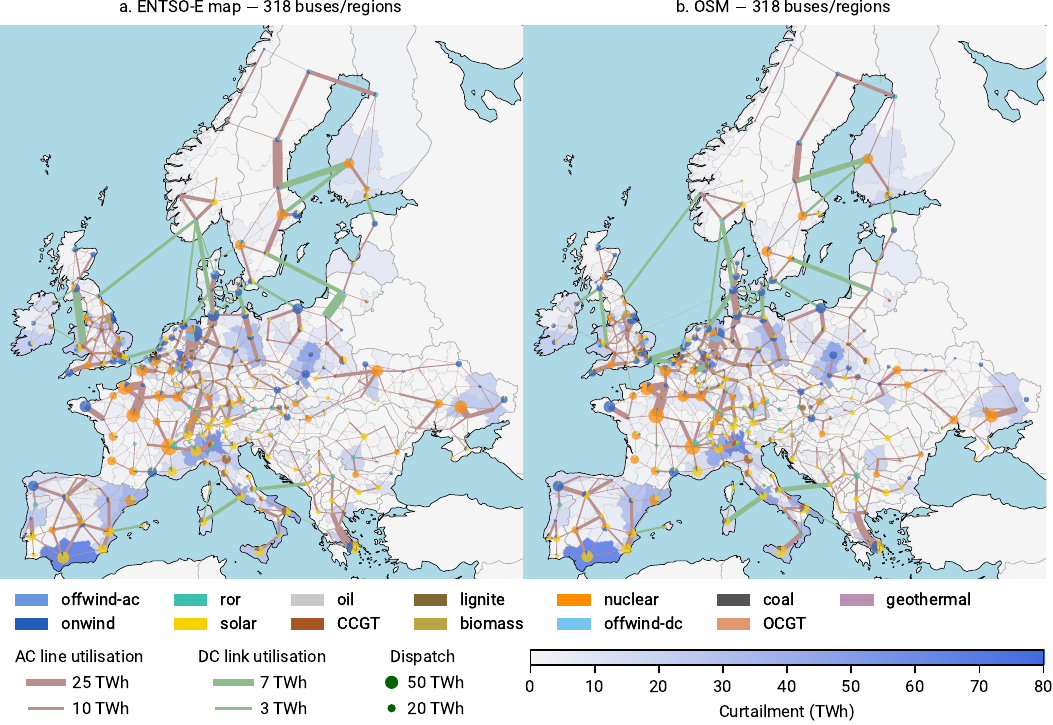}
    \caption{Regional dispatch, line utilisation and curtailment. \textit{A map comparing nominal ratings of the two clustered grids is provided in the Figure \ref{fig:results_sidebyside_capacity}.}}
    \label{fig:results_sidebyside}
\end{figure}

\begin{figure}[!htbp]
    \centering
    \includegraphics{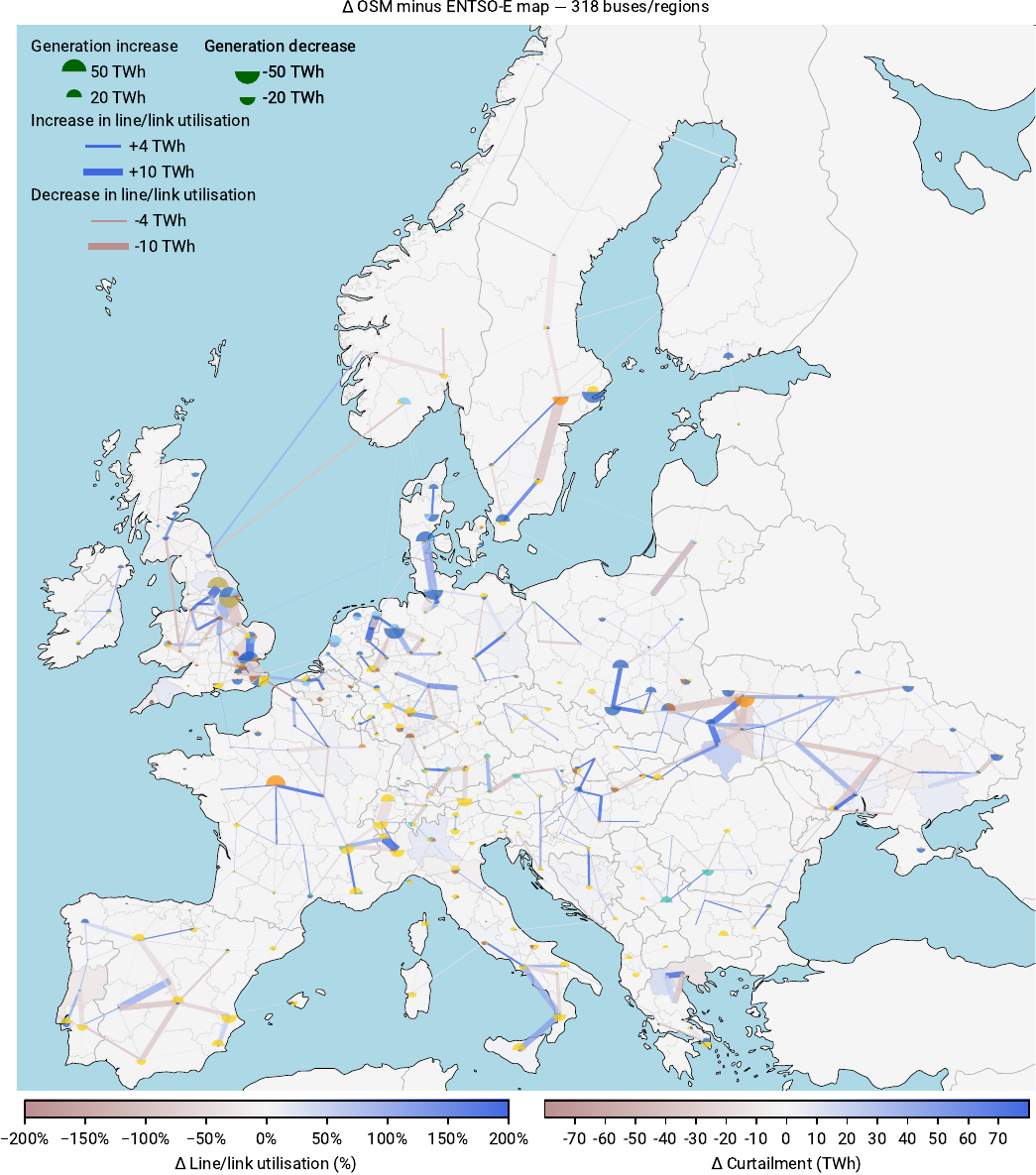}
    \caption{Regional dispatch, line utilisation and curtailment. \textit{Blue indicates an increase in curtailment or line utilisation from the \acrshort{entsoe} map to the \gls{osm}-based transmission grid, while red indicates a decrease. For full transparency, note that this map shows an outer join of all transmission grid elements, including lines and links that are not present in the other network.}}
    \label{fig:results_delta}
\end{figure}

\begin{figure}[!htbp]
    \centering
    \includegraphics{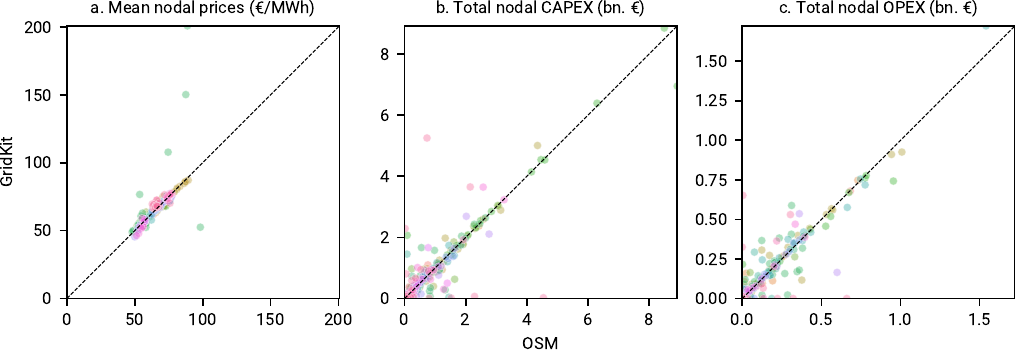}
    \caption{Comparison of nodal prices, capital expenditures (CAPEX) and operational expenditures (OPEX). \textit{Outliers nodal prices (a): Great-Britain (green). Outliers nodal CAPEX (b): Ukraine (pink).}}
    \label{fig:results_cost_scatter}
\end{figure}

We solve the two optimisation problems on a high-performance cluster (AMD EPYC 7543 32-Core processor) using up to \SI{130}{\giga\byte} of memory. Each problem takes up to 274 iterations to converge, translating into more than 18 hours. Running the model for the two transmission grid representations, we find that regional results align closely, including the dispatch of generation assets, utilisation of lines and curtailment (Figure \ref{fig:results_sidebyside}). This is also true for the aggregated picture (Table \ref{tab:results}). Total system costs drop from 312.94 bn. \euro{} in the \acrshort{entsoe} map based run to 311.7 bn. € in the \acrshort{osm}-based optimisation, corresponding to mere \SI{0.40}{\percent} in difference. The higher weighted degree for both the base and clustered \acrshort{osm}-based high-voltage grid (Figure \ref{fig:hist_degree_nuts2}) indicate a higher topological connectivity, potentially translating into higher degrees of freedom in the optimisation problem compared to the \acrshort{entsoe} map based run. This is confirmed when we look into i) the line and link utilisation and ii) compare the investments. Since aggregate statistics over a continental area can be deceiving (smooth out errors), we show the differences in regional generation, line and link utilisation as well as curtailment in Figure \ref{fig:results_delta}. Here, we can clearly see that the high-voltage grid in \acrshort{osm} is higher utilised than its \acrshort{entsoe} map based counterpart. We can also observe that the \acrshort{entsoe} map based model compensates by investing more into generation (especially decentral, semi-circles in the bottom half) and storage capacities, i.e. +\SI{4.9}{\giga\watt} in solar photovoltaics, +\SI{4.8}{\giga\watt} in onshore wind, +\SI{1.1}{\giga\watt} in offshore wind (\acrshort{ac}), and +\SI{2.4}{\giga\watt} in battery storage. In the \acrshort{osm}-based model run, we see a slightly stronger build-out of \acrshort{dc} connected offshore wind (+\SI{2.7}{\giga\watt}). We also provide an overview of average electricity prices, CAPEX and OPEX at nodal level in Figure \ref{fig:results_cost_scatter}.

\begin{table}[!htbp]
    \centering
    \begin{filecontents*}{datatables/tab_results.csv}
,System costs (bn. €),CAPEX (bn. €),OPEX (bn. €),Curtailment (TWh),Generation (TWh),Line volume base (MVAkm),Line volume clustered (MVAkm)
GridKit,312.94,274.46,38.48,2177.91,3120.0,410.89,431.47
OSM,311.7,273.47,38.24,2175.57,3110.03,424.45,424.94
Delta (Percent),-0.4,-0.36,-0.64,-0.11,-0.32,3.3,-1.51
\end{filecontents*}
\csvreader[
        tabular=|r|c|c|c|c|c|,
        table head=\hline & 
        \makecell{\textbf{System costs} \\ (bn. \euro{}/a)} &
        \makecell{\textbf{CAPEX} \\ (bn. \euro{}/a)} &
        \makecell{\textbf{OPEX} \\ (bn. \euro{})/a} &
        \makecell{\textbf{Curtailment} \\ (\SI{}{\tera\watt\hour}/a)} &
        \makecell{\textbf{Generation} \\ (\SI{}{\tera\watt\hour}/a)}
        \\\hline,
        late after line=\\\hline,
        head to column names
    ]{datatables/tab_results.csv}{}%
    { 
        \ifthenelse{\equal{\thecsvrow}{1}}{ENTSO-E map}{%
        \ifthenelse{\equal{\thecsvrow}{2}}{OSM}{%
        \ifthenelse{\equal{\thecsvrow}{3}}{ Delta (\SI{}{\percent})}{}}}& 
        \csvcolii & \csvcoliii & \csvcoliv & \csvcolv & \csvcolvi
    }
    \caption{Comparison of key result metrics between \acrshort{entsoe} map and \acrshort{osm}-based transmission grid.}
    \label{tab:results}
\end{table}

Bottlenecks in both model runs are located in the same regions, contributing to an annual curtailment in the range of \SI{2176}{\tera\watt\hour} to \SI{2178}{\tera\watt\hour}. More prominent differences in line utilisation are visible in Norway and Poland from North to South, in the western region of Ukraine, southern and central Spain around the Madrid area, as well as southern Italy. 

Overall, the results of the two model runs are very similar, indicating that i) both grids seem to adequately represent reality and ii) the \gls{osm}-based transmission grid is a suitable replacement for the \acrshort{entsoe} map based grid. The higher utilisation of the \gls{osm}-based transmission grid is in line with the higher topological connectivity of the network. The differences in the investment decisions are marginal and can be attributed to the differences in grid topology.

We have shown that the dataset is in good agreement with official statistics and the \acrshort{entsoe} map, and the results of a representative PyPSA-Eur model instance based on the two high-voltage grid datasets are very similar. Its core strengths lie in the high level of geographic detail and the continuous updates to the \gls{osm} database in combination with a strong PyPSA-Eur user base. The workflow is completely transparent and the data is provided openly. While we cannot guarantee the correctness of the data, as only \acrshortpl{tso} have access to the real grid data, we believe that the dataset provides the best publicly available representation of the European high-voltage grid.

\newpage

\newpage

\section*{Usage Notes}
\pdfbookmark[1]{Usage Notes}{usage-notes}
The published dataset is provided under the \gls{odbl} 1.0 licence. Geoinformation is encoded in the WGS84 (EPSG:4326) coordinate system. Although the dataset and workflow are provided as part of PyPSA-Eur, they are also suitable for a wide range of other applications, including network analyses, power flow calculations, as well as an input for other energy energy system models and frameworks. Note that further validation and testing is needed for purposes outside the original scope of this work. 

\begin{itemize}
    \item In order to reproduce the network with the latest \gls{osm} data, the configuration file `config.yaml': needs to be set to \colorcode{base_network = `osm-prebuilt'} and the command \colorcode{snakemake base_network -call} needs to be executed.
    \item To rebuild the network from scratch, this setting can be changed to \colorcode{base_network = `osm-raw'}, followed by the command \colorcode{snakemake prepare_osm_network_release -call}. 
    \item Per default, buses within the perimeter of a \SI{5000}{\meter} radius are merged together. This value can be changed in the script, however this may change the topological connectedness of the obtained network. 
    \item Networks can also be built for specific countries, regions or a subset of the countries within PyPSA-Eur by setting list of \colorcode{countries} the configuration file.
\end{itemize}
To abide to the fair use policy of the \acrshort{osm} Overpass turbo \acrshort{api}, we kindly encourage users to download the prebuilt network topology from the Zenodo repository and only rebuild the network, if necessary. We would further like to encourage readers to actively contribute to the \acrshort{osm} database.

\section*{Code availability}
\pdfbookmark[1]{Code availability}{code-availability}
The code to replicate the entire workflow and dataset is provided as part of PyPSA-Eur and released as free software under the MIT licence. Different licences and terms of use may apply to the underlying input data.
\begin{itemize}
    \item PyPSA-Eur \autocite{horschPyPSAEurOpenOptimisation2018} on GitHub: \\ \href{https://github.com/pypsa/pypsa-eur}{https://github.com/pypsa/pypsa-eur}
    \item Version 0.3 of the prebuilt network\autocite{xiongPrebuiltElectricityNetwork2024} based on \acrshort{osm} data can be retrieved via the Zenodo repository. This link will also point to future updates: \\ \href{https://zenodo.org/records/13358976}{https://zenodo.org/records/13358976}
\end{itemize}

\phantomsection
\bibliography{osm-for-pypsa-eur}

\section*{Acknowledgements} 
\pdfbookmark[1]{Acknowledgements}{acknowledgements}
The authors would like to thank the \gls{osm} community for their tremendous motivation and efforts in providing the data underlying this work. Further, the authors would like to thank the PyPSA-Earth community for setting the groundwork and initiating the move to grid data based on \gls{osm}. Last but not least, the authors would like thank Ekaterina Fedotova and Emmanuel Bolarinwa for early substantial developments and validations in integrating \acrshort{osm} into PyPSA-Earth and Philipp Glaum for technical sparring and support. Map data copyrighted OpenStreetMap contributors and available from \href{https://www.openstreetmap.org}{https://www.openstreetmap.org}. 

This work was supported by the German Federal Ministry for Economic Affairs and Climate Action (BMWK) under Grant No. 03EI4083A (RESILIENT) and Italian Ministry of University and Research (MUR) CUP I53C23002650007 (RESILIENT). This project has been funded by partners of the CETPartnership (https://cetpartnership.eu/) through the Joint Call 2022. As such, this project has received funding from the European Union's Horizon Europe research and innovation programme under grant agreement no. 101069750.

\section*{Author contributions statement}
\pdfbookmark[1]{Author contributions statement}{author-contributions-statement}
B.X. -- Conceptualisation, Data curation, Methodology, Software/Programming, Model building and validation, Visualisation, Writing -- original draft, review and editing. 
D.F. -- Conceptualisation, Software/Programming, Writing -- review and editing.
F.N. -- Conceptualisation, Discussion, Writing -- review and editing.
I.R. -- Discussion, Writing -- review and editing.
T.B. -- Supervision, Writing -- review and editing.

\section*{Competing interests} 
\pdfbookmark[1]{Competing interests}{competing-interests}
The authors declare no competing interests.

\newpage
\section*{Figures \& Tables}
\pdfbookmark[1]{Figures \& Tables}{figures-and-tables}

\begin{table}[!htbp]
    \centering
    \begin{tabular}{|p{0.25\textwidth}|p{0.71\textwidth}|}
    \hline
    \textbf{Grid element} & \textbf{Overpass turbo query}  \\
    \hline
    \acrshort{ac} power lines and cables & \colorcode{way[`power'=`line']} and \colorcode{way[`power'=`cable']} \\
    \hline
    \acrshort{dc} links & \colorcode{relation[`route'=`power'][`frequency'=`0']} \\
    \hline
    substations & \colorcode{way[`power'=`substation']} and \colorcode{relation[`power'=`substation']} \\
    \hline
    \end{tabular}
    \caption{Overpass turbo queries used to extract the high-voltage grid elements from \acrshort{osm}.}
    \label{tab:overpass_queries}
\end{table}

\begin{table}[!htbp]
    \centering
    \begin{tabular}{|p{2.5cm}|p{2.6cm}|p{3.5cm}|p{3cm}|p{1.5cm}|p{2cm}|}
    \hline
    \textbf{Project} & \textbf{Regional scope} & \textbf{Tools/data} & \textbf{Last update} & \textbf{Geo\-referenced} & \textbf{Data \newline published} \\
    \hline
    50Hertz static grid model\autocite{50hertzStaticGridModel2022} & Germany (50Hertz control area) & based on asset inventory & April 2022 \newline (updated once a year) & Yes & Yes \\
    \hline
    ELMOD\autocite{egererElectricitySectorData2014} & Europe & ENTSO-E interactive map\autocite{entso-eENTSOETransmissionSystem} plus manual changes & January 2014 \newline (data not published) & Yes & No \\
    \hline
    ELMOD-DE\autocite{egererOpenSourceElectricity2016} & Germany & VDE, \acrshort{tso} maps, and \acrshort{osm} & March 2016
     \newline (single release) & Yes & Yes (not reproducible) \\
    \hline
    Hutcheon \& Bialek\autocite{hutcheonUpdatedValidatedPower2013} & Europe (UCTE plus Balkan region) & PowerWorld model & June 2013 \newline (updated once) &  No & Yes (not reproducible) \\
    \hline
    JAO static grid model\autocite{jaoStaticGridModel2023} & CORE capacity calculation region & based on CORE \acrshort{tso} asset inventory & April 2024 \newline (updated frequently) &  No & Yes \\
    \hline
    PyPSA-Eur\autocite{horschPyPSAEurOpenOptimisation2018} & Europe & ENTSO-E interactive map\autocite{entso-eENTSOETransmissionSystem} using GridKit\autocite{wiegmansGridKitExtractENTSOE2016} plus manual changes& January 2022 \newline (updated once) &  Yes & Yes (reproduction complex) \\
    \hline
    osmTGmod \autocite{OsmTGmodDocumentation2017} & Germany & \acrshort{osm} using Osmosis (SQL and Java)\autocite{wiegmansGridKitExtractENTSOE2016} & November 2017 \newline (single release) &  Yes & Yes (reproduction complex)\\
    \hline
    SciGrid (Power) \autocite{medjroubiOpenDataPower2017} & Europe, Germany & \acrshort{osm} using GridKit\autocite{wiegmansGridKitExtractENTSOE2016} & November 2015 \newline (updated once) & Yes & Yes (reproduction complex)\\
    \hline
    \bfseries This publication & Europe & \acrshort{osm} using Overpass turbo \acrshort{api} and Python & August 2024 \newline (updated frequently) &  Yes & Yes \autocite{xiongPrebuiltElectricityNetwork2024} (reproducible)\\
    \hline
    \end{tabular}
    \caption{Notable projects and datasets modelling the high-voltage grid in Europe (alphabetical order). \textit{Note that the specific regional scope referring to `Europe' may vary across the listed projects. A comparison of our dataset with the 50Hertz static grid model is shown in Figure \ref{fig:osm_vs_50Hertz_sgm}.}}
    \label{tab:network_projects}
\end{table}

\newpage
\begin{table}[!htbp]
    \centering
    \begin{filecontents*}{datatables/tab_dc_links.csv}
url,name,from,to,voltage,p_nom,length
\href{https://www.openstreetmap.org/relation/8193755}{relation/8193755},Aachen Lüttich Electricity Grid Overlay (ALEGrO),BE,DE,320,1000,90
\href{https://www.openstreetmap.org/relation/3918230}{relation/3918230},Baltic Cable,DE,SE,450,600,260
\href{https://www.openstreetmap.org/relation/14126301}{relation/14126301},BritNed,GB-ENG,NL,450,1000,254
\href{https://www.openstreetmap.org/relation/8099179}{relation/8099179},Caithness-Moray Link,GB-SCT,GB-SCT,320,1200,136
\href{https://www.openstreetmap.org/relation/17631956}{relation/17631956},Conexión Mediterránea Transporte Alta tensión (COMETA),ES,ES,250,400,238
\href{https://www.openstreetmap.org/relation/8185420}{relation/8185420},Copenhagen-Brussels-Amsterdam cable (COBRAcable),DK,NL,320,700,328
\href{https://www.openstreetmap.org/relation/2127794}{relation/2127794},Cross-Channel (IFA 2000),FR,GB-ENG,270,2000,70
\href{https://www.openstreetmap.org/relation/8184641}{relation/8184641},East-West Interconnector,IE,GB-WLS,400,500,257
\href{https://www.openstreetmap.org/relation/15772117}{relation/15772117},ElecLink,FR,GB-ENG,320,1000,53
\href{https://www.openstreetmap.org/relation/8184630}{relation/8184630},Estlink 1,EE,FI,150,350,105
\href{https://www.openstreetmap.org/relation/6886400}{relation/6886400},Estlink 2,EE,FI,450,650,175
\href{https://www.openstreetmap.org/relation/8184631}{relation/8184631},Fenno-Skan,FI,SE,400,500,229
\href{https://www.openstreetmap.org/relation/3391954}{relation/3391954},Fenno-Skan 2,FI,SE,500,800,305
\href{https://www.openstreetmap.org/relation/3392001}{relation/3392001},Gotlandslänken (Gotland 2+3),SE,SE,150,260,96
\href{https://www.openstreetmap.org/relation/5487095}{relation/5487095},Great Belt power link,DK,DK,400,600,56
\href{https://www.openstreetmap.org/relation/9934065}{relation/9934065}; \href{https://www.openstreetmap.org/relation/9934066}{relation/9934066},Interconnexion Electrique France-Espagne (INELFE),ES,FR,320,2000,64
\href{https://www.openstreetmap.org/relation/10377412}{relation/10377412},Interconnexion France-Angleterre 2 (IFA-2),FR,GB-ENG,320,1000,231
\href{https://www.openstreetmap.org/relation/8185664}{relation/8185664},Italy-Greece,GR,IT,400,500,302
\href{https://www.openstreetmap.org/relation/9982798}{relation/9982798},Italy–Corsica–Sardinia (SACOI) - Corsica-Sardinia,FR,IT,200,300,141
\href{https://www.openstreetmap.org/relation/3391794}{relation/3391794},Italy–Corsica–Sardinia (SACOI) - Italy-Corsica,FR,IT,200,300,241
\href{https://www.openstreetmap.org/relation/2505320}{relation/2505320},KONTEK,DE,DK,400,600,166
\href{https://www.openstreetmap.org/relation/8184629}{relation/8184629},Konti-Skan,DK,SE,300,680,142
\href{https://www.openstreetmap.org/relation/8185767}{relation/8185767},Montenegro-Italy (MON.ITA),IT,ME,500,1200,412
\href{https://www.openstreetmap.org/relation/6914309}{relation/6914309},Moyle interconnector,GB-NIR,GB-SCT,500,500,62
\href{https://www.openstreetmap.org/relation/8185487}{relation/8185487},Nemo Link,BE,GB-ENG,400,1000,139
\href{https://www.openstreetmap.org/relation/9965201}{relation/9965201},NorNed,NL,NO,450,700,578
\href{https://www.openstreetmap.org/relation/8185455}{relation/8185455},NordBalt,LT,SE,300,700,448
\href{https://www.openstreetmap.org/relation/16213216}{relation/16213216},NordLink,DE,NO,525,1400,620
\href{https://www.openstreetmap.org/relation/13295785}{relation/13295785},North Sea Link,GB-ENG,NO,515,1400,723
\href{https://www.openstreetmap.org/relation/8185711}{relation/8185711},Sardegna-Penisola Italiana (SAPEI),IT,IT,500,1000,417
\href{https://www.openstreetmap.org/relation/3391931}{relation/3391931},Skagerrak (1-3),DK,NO,350,940,241
\href{https://www.openstreetmap.org/relation/8184632}{relation/8184632},Skagerrak (4),DK,NO,500,700,226
\href{https://www.openstreetmap.org/relation/3392010}{relation/3392010},SwePol,PL,SE,450,600,255
\href{https://www.openstreetmap.org/relation/8184633}{relation/8184633},SydVästlänken,NO,SE,300,1200,253
\href{https://www.openstreetmap.org/relation/15781671}{relation/15781671},Viking Link,DK,GB-ENG,525,1400,742
\href{https://www.openstreetmap.org/relation/15775538}{relation/15775538},Western HVDC Link,GB-SCT,GB-WLS,600,2250,401
\end{filecontents*}
\csvreader[
        tabular=|C{2.5cm}|C{4.8cm}|C{1.4cm}|C{1.4cm}|C{1.2cm}|C{1.2cm}|C{1.9cm}|, 
        table head=\hline 
        \bfseries \makecell{\acrshort{osm} relation \\ identifier} & 
        \makecell{\bfseries \acrshort{dc} project name \\ (sorted alphabetically)} & 
        \bfseries From & 
        \bfseries To & 
        \makecell{\textbf{Voltage} \\(\SI{}{\kilo\volt})} & 
        \makecell{\textbf{Rating} \\(\SI{}{\mega\watt})} & 
        \makecell{\textbf{Calc. length} \\(\SI{}{\kilo\meter})} 
        \\\hline,
        late after line=\\\hline,
        head to column names
    ]{datatables/tab_dc_links.csv}{}%
    {\csvcoli & \csvcolii & \csvcoliii & \csvcoliv & \csvcolv & \csvcolvi & \csvcolvii}

    \caption{List of \acrshort{dc} projects in the \acrshort{osm}-based transmission grid. \textit{Note that \acrshort{osm} relation identifiers are unique and persistent as long as the object is not deleted. Projects can be directly accessed via the \acrshort{osm} website by clicking on their respective relation identifier in the table.}}
    \label{tab:dc_links}
\end{table}

\newpage
\begin{table}[!htbp]
    \centering 
    \begin{filecontents*}{datatables/tab_s_nom.csv}
Voltage,Line type,Nominal current,Circuits,Total rating,Length
220,Al/St 240/40 2-bundle 220.0,1.290,1,492,52897
220,Al/St 240/40 2-bundle 220.0,1.290,2,983,22719
220,Al/St 240/40 2-bundle 220.0,1.290,3,1475,524
220,Al/St 240/40 2-bundle 220.0,1.290,4,1966,37
225,Al/St 240/40 2-bundle 220.0,1.290,1,503,19355
225,Al/St 240/40 2-bundle 220.0,1.290,2,1005,4465
236,Al/St 240/40 2-bundle 220.0,1.290,1,527,19
275,Al/St 240/40 3-bundle 300.0,1.935,1,922,1097
275,Al/St 240/40 3-bundle 300.0,1.935,2,1843,2845
300,Al/St 240/40 3-bundle 300.0,1.935,1,1005,4127
300,Al/St 240/40 3-bundle 300.0,1.935,2,2011,20
330,Al/St 240/40 3-bundle 300.0,1.935,1,1106,17335
330,Al/St 240/40 3-bundle 300.0,1.935,2,2212,1115
380,Al/St 240/40 4-bundle 380.0,2.580,1,1698,13971
380,Al/St 240/40 4-bundle 380.0,2.580,2,3396,13989
380,Al/St 240/40 4-bundle 380.0,2.580,3,5094,362
380,Al/St 240/40 4-bundle 380.0,2.580,4,6792,259
400,Al/St 240/40 4-bundle 380.0,2.580,1,1787,56750
400,Al/St 240/40 4-bundle 380.0,2.580,2,3575,30977
400,Al/St 240/40 4-bundle 380.0,2.580,4,7150,84
412,Al/St 240/40 4-bundle 380.0,2.580,1,1841,32
420,Al/St 240/40 4-bundle 380.0,2.580,1,1877,4839
420,Al/St 240/40 4-bundle 380.0,2.580,3,5631,11
500,Al/St 240/40 4-bundle 380.0,2.580,1,2234,248
750,Al/St 560/50 4-bundle 750.0,4.160,1,5404,4148
\end{filecontents*}
\csvreader[
        tabular=|c|c|r|r|r|r|,
        table head=\hline\bfseries $\bm{U_{nom}^{OSM}}$ (\SI{}{\kilo\volt}) & \bfseries Line type & \bfseries $\bm{I_{nom}^
        {pandapower}}$ (\SI{}{\ampere}) & \bfseries $\bm{n_{circuits}}$ & \bfseries $\bm{S_{nom}^{AC}}$ (\SI{}{\mega\volt\ampere}) & \bfseries Total route length (\SI{}{\kilo\meter}) \\\hline,
        late after line=\\\hline,
        head to column names
    ]{datatables/tab_s_nom.csv}{}%
    {\csvcoli & \csvcolii & \csvcoliii & \csvcoliv & \csvcolv & \csvcolvi}
    \caption{\label{tab:s_nom} Nominal capacities in the \acrshort{osm} base network  \acrshort{ac} lines and cables to pandapower standard type library.}
\end{table}

\begin{figure}[!htbp]
    \centering
    \includegraphics{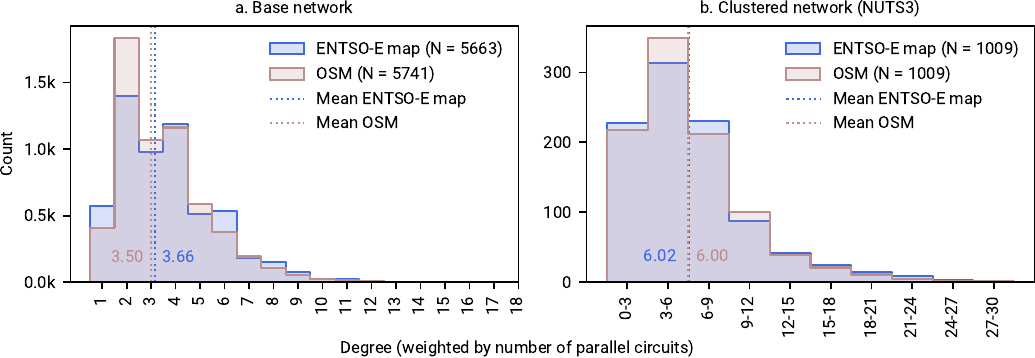}
    \caption{Comparison of the weighted degree distribution in both networks before and after clustering (NUTS3). \textit{Ukraine at geoBoundaries\autocite{runfolaGeoBoundariesGlobalDatabase2020} administration level 1, Moldova in full bus resolution.}}
    \label{fig:hist_degree_nuts3}
\end{figure}

\newpage
\begin{figure}[!htbp]
    \centering
    \includegraphics{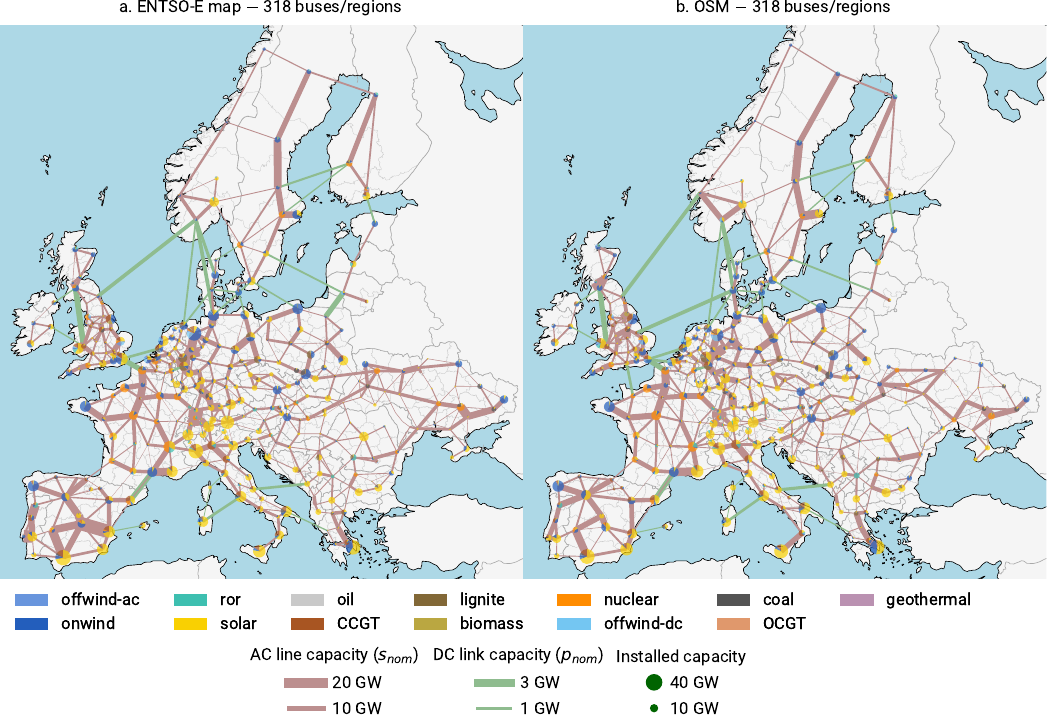}
    \caption{Clustered \acrshort{ac} line capacities, \acrshort{dc} link nominal ratings, and optimal generation capacities.}
    \label{fig:results_sidebyside_capacity}
\end{figure}

\newpage
\begin{figure}[!htbp]
    \centering
    \includegraphics{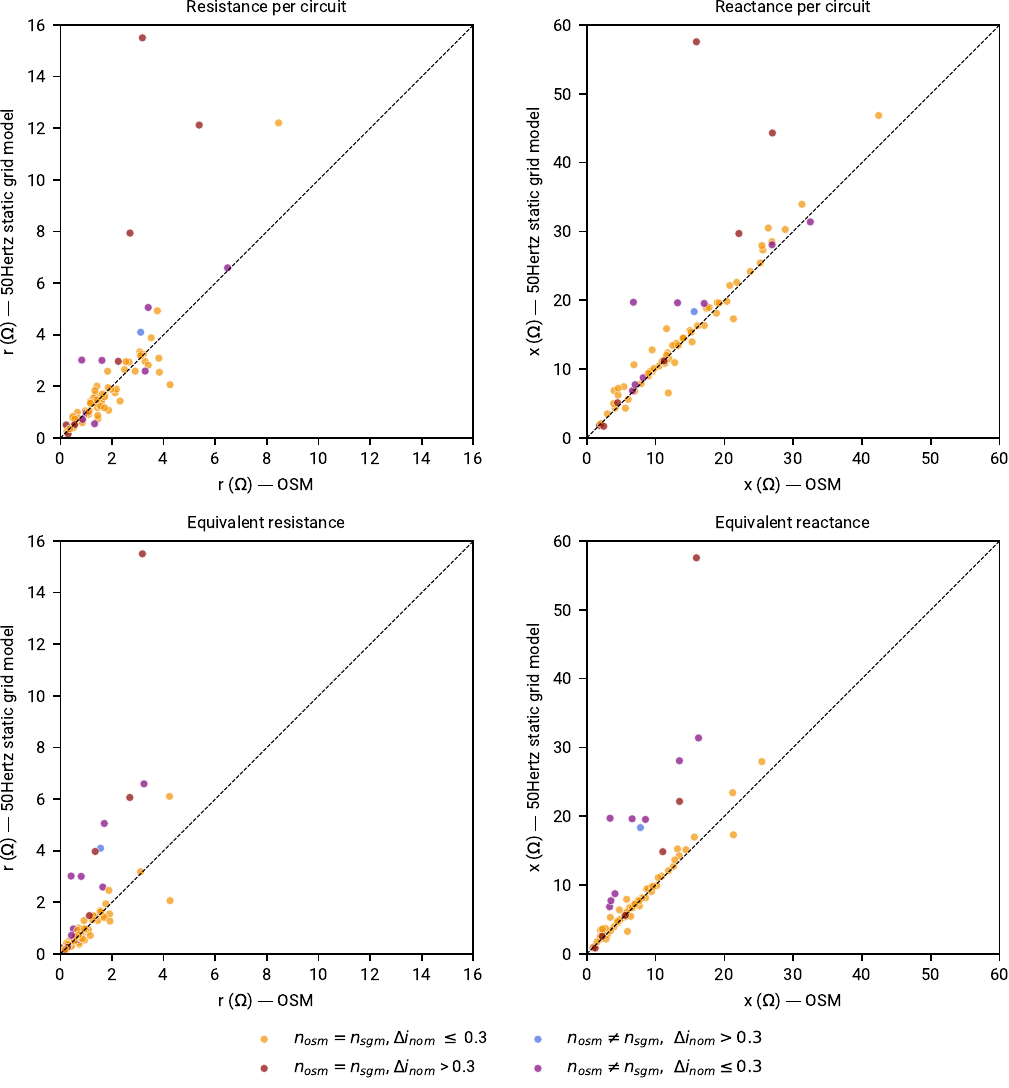}
    \caption{Comparison of \acrshort{ac} line/cable resistance and reactance between the \acrshort{osm}-based transmission grid and reference 50Hertz static grid model.\autocite{50hertzStaticGridModel2022} \textit{$n_{osm}$ and $n_{sgm}$ refer to the number of parallel circuits for a distinct line in each network, while $\Delta i_{nom} = \frac{|i_{nom, osm} - i_{nom, sgm}|}{i_{nom,sgm}}$ refers to the relative change in underlying nominal current.}}
    \label{fig:scatter_reactance_resistance}
\end{figure}

Figure \ref{fig:scatter_reactance_resistance} was generated by mapping \acrshort{ac} lines and cables of the \acrshort{osm}-based transmission grid to the 50Hertz static grid model (SGM) using \acrshort{osm} tags and SGM names (right join). Note that this data explains \SI{4475}{\kilo\meter} of \SI{5126}{\kilo\meter} in route length, as not all lines could mapped. For \SI{79}{\percent} of the data, using pandapower's standard line types\autocite{thurnerPandapowerOpenSourcePython2018} for calculating the resistance and reactance comes close to official data in the SGM (orange). Purple data points a discrepance primarily due to unequal number of parallel circuits in both datasets (SGM data larger by factor 2). Red and blue data points indicate that underlying line types are entirely different. This is the case for some lines where SGM e.g. has a newer \SI{380}{\kilo\volt} (allowing higher higher currents) or weaker \SI{220}{\kilo\volt} line type.

\end{document}